\documentclass{article}
\usepackage[final]{nips_2017}
\usepackage[utf8]{inputenc}
\usepackage[T1]{fontenc}   
\usepackage{hyperref,url,amsfonts,nicefrac,microtype}      
\usepackage{booktabs,multirow,hhline,graphicx,arydshln, subfig}
\usepackage{algorithm}
\usepackage{algorithmic}
\usepackage{tabularx}
\usepackage[export]{adjustbox}
\PassOptionsToPackage{square,numbers}{natbib}
\bibpunct{[}{]}{,}{n}{}{;}

\makeatletter
\def\adl@drawiv#1#2#3{%
        \hskip.5\tabcolsep
        \xleaders#3{#2.5\@tempdimb #1{1}#2.5\@tempdimb}%
                #2\z@ plus1fil minus1fil\relax
        \hskip.5\tabcolsep}
\newcommand{\cdashlinelr}[1]{%
  \noalign{\vskip\aboverulesep
           \global\let\@dashdrawstore\adl@draw
           \global\let\adl@draw\adl@drawiv}
  \cdashline{#1}
  \noalign{\global\let\adl@draw\@dashdrawstore
           \vskip\belowrulesep}}
\makeatother

\title{Wrist Sensor Fusion Enables Robust Gait Quantification Across Walking Scenarios}
\author{
  Zeev Waks, Itzik Mazeh, Chen Admati, Michal Afek, Yonatan Dolan, Avishai Wagner\\
  Intel - Advanced Analytics\\
  \texttt{\{zeev.waks, itzik.mazeh, chen.admati, michal.afek,}\\
  \texttt{yonatan.dolan, avishai.wagner\}@intel.com}\\
  }

\begin{document}
\maketitle
\begin{abstract}
Quantifying step abundance via single wrist-worn accelerometers is a common approach for encouraging active lifestyle and tracking disease status. Nonetheless, step counting accuracy can be hampered by fluctuations in walking pace or demeanor. Here, we assess whether the use of various sensor fusion techniques, each combining bilateral wrist accelerometer data, may increase step count robustness. By collecting data from 27 healthy subjects, we find that high-level step fusion leads to substantially improved accuracy across diverse walking scenarios. Gait cycle analysis illustrates that wrist devices can recurrently detect steps proximal to toe-off events. Collectively, our study suggests that dual-wrist sensor fusion may enable robust gait quantification in free-living environments. 
\end{abstract}

\section{Introduction}
The growth of the quantified self movement has sparked substantial interest in measuring the extent of physical activity an individual performs. One widespread approach for tracking daily activity is by counting walking steps. This is often accomplished using wrist-worn pedometers containing triaxial accelerometers. Beyond the public health value of increased ambulation, step counting interventions yield modest weight loss \cite{richardson2008meta}, and may aid in diabetes, cardiovascular disorders, and COPD care \cite{bassett2017step}.

Despite their prevalence, commercial wrist-worn pedometers often display markedly inferior accuracy compared to their counterparts located on other body parts, for instance the waist or ankles \cite{storm2015step,husted2017accuracy,chow2017accuracy}. Indeed, wrist-based step counting does not appear robust to walking pace and other walking style variations, with double digit error rates being fairly common \cite{storm2015step,chow2017accuracy,chen2016accuracy}. In fact, it has been argued that step counts derived using accelerometers worn on the waist and wrist are generally not comparable under both laboratory and free-living conditions \cite{tudor2015comparison}, with waist placement being superior \cite{chow2017accuracy}. 

Body-worn micro electromechanical system (MEMS) inertial sensors including accelerometers and gyroscopes can also be used to quantify gait abnormalities in movement disorders. This is primarily achieved in controlled settings by detecting specific gait phases, especially the heel strike (initial contact) and toe-off (terminal contact) events. For example, inertial sensors have been used to track cadence inconsistencies and axial symptoms such as step time asymmetry in patient cohorts \cite{del2016free,mirelman2016arm}. 

In free-living environments, wrists would presumably serve as attractive sites for monitoring gait pathologies given the wide adoption of wrist-worn devices. However, due to arm movement, use of individual wrist-worn sensors for measuring lower-limb, time-dependent aspects such as initial foot contact and terminal contact is challenging. We hypothesized that concurrent use of opposing wrist sensors, via sensor fusion, may help overcome the limitations of single-wrist devices. 

Sensor fusion refers to the combining of sensory data such that resulting information may be better than that from individual sources (reviewed in \cite{elmenreich2002sensor}). Algorithms used for synchronous multi-sensor fusion include, among others, averaging, Borda count voting, fuzzy logic, Kalman filters, and inference methods. Sensor fusion architectures generally fall under three categories: low-level (raw data), intermediate-level (features), or high-level (combining decisions, in this paper detected steps).

Perhaps surprisingly, there are limited examples of sensor fusion using body-worn inertial sensors. Indeed, multi-sensor studies often compare the outputs of individual sensors rather than perform fusion to improve a measure of interest. Existing fusion examples typically focus on activity classification via intermediate-level fusion, for example by classifying daily activities \cite{gao2014evaluation} or detecting falls \cite{lopez2017analyzing}. Sensor fusion for gait quantification is scarce, with one exception using ankle, thigh, and waist sensors, but not wrist devices, and focusing on a single high-level fusion approach \cite{fortune2015step}.

Given the lack of precedent, methodical comparison of multiple fusion approaches for gait quantification would be of high value, especially using wrist sensors. The present study explores this possibility by comparing sensor fusion with single sensor results. Using eight types of walking tasks, we assess whether dual-wrist sensor fusion can facilitate robust step counting and gait phase detection. 

\section{Data collection and labels}
Data was collected from 27 healthy volunteers (18 male, 9 female) aged between 18 and 50. Subjects received instructions and subsequently performed eight separate walking tasks that simulate multiple walking types (Table \ref{T:setting2}). A total of N=214 tasks were successfully collected. Tasks were performed while wearing six synchronized, wireless devices, each containing a 128Hz triaxial accelerometer and 128Hz triaxial gyroscope (Opal sensors, APDM). The six devices were placed on both ankles, both wrists, the lumbar spine, and the trunk, per manufacturer configuration. The entire process including instructions, placement, and removal of the devices was performed in less than 30 minutes. 

Each walking task consisted of a 74.6m uninterrupted walk in a flat, rectangular course (33.5m X 3.8m) with a mean duration of 64.3s [range 30.4s-110.6s]. Subjects self-counted their steps, with a mean step count of 104.4 steps [range 67-151]. These counts were not used for training or evaluation.

Ankle gyroscope signals were used to derive step count labels and to determine heel strike and toe-off timings \cite{salarian2004gait,aminian2002spatio}. The method, implemented by the device manufacturer (APDM), is considered highly reliable and has been previously used as a label \cite{storm2015step}. Gait events were identified by matching a template, specifically a double pendulum model for leg swing and an inverse double pendulum for stance (foot on ground). Segments analyzed by the manufacturer software, accounting for the large majority of gait cycles, were used for step count evaluation. To further ensure accuracy, we discarded the 5\% of samples with the largest cadence difference between subject step counts and ankle-derived values. This increased the correlation between the measures from r=0.81 (N=214) to r=0.96 (N=203).

\begin{table}[hb]
  \caption{Data collection walking tasks. Sample sizes shown after outlier removal.}
  \label{T:setting2}
  \centering
  \begin{tabular}{ll}
  \toprule
  \textbf{Category} & \textbf{Walking type (N)}\\
  \hhline{--}
  Unconstrained & Slow pace (25), Comfortable pace (26), Fast pace (24)\\
  \cdashlinelr{1-2}
  \multirow{2}{*}{Arms constrained} & Holding bag in right hand (27), Holding cellphone with two hands (26),\\
  & Hands not swinging alongside body (27)\\
  \cdashlinelr{1-2}
  Asymmetrical & Without right shoe (25), pretending to use cane in right hand (23)\\
  \hhline{--}
  \textbf{Total} & \textbf{203}\\
  \bottomrule
  \end{tabular}
\end{table}

\section{Methods}
This work evaluates no fusion (single side), low-level fusion, and high-level fusion step detection approaches. All approaches first converted axis-level raw data into signal magnitude (norm) \textbf{$\sqrt[]{x_t^2+y_t^2+z_t^2}$}, followed by smoothing with a centered moving average. Peak detection was used for step counting as previously recommended for non-fixed accelerometers given its high performance \cite{brajdic2013walk}. Peaks were detected using first-order difference with parameters for minimum peak height and minimum window between adjacent peaks \cite{peakutils}. In the case of multiple peaks within the minimum window, the highest peak was chosen. Each peak was considered a single step. All parameters were tuned by five-fold cross validation using all N=203 samples by minimizing the root mean square error (values shown in Table \ref{T:params}). This resulted in negligible smoothing for a few of the approaches.

\begin{table}[ht]
  \caption{Mean parameter values of the six algorithms using five-fold cross validation. Minimum peak amplitude values are for signal values $\in$ [0, 1], min-max normalized using data from all samples.}
  \label{T:params}
  \centering
  \begin{tabular}{lcccccc}
    \toprule
    \multirow{3}{*}{\textbf{Parameter}} & \multicolumn{6}{c}{\textbf{Algorithm}}\\
    \cmidrule(lr){2-7}
    & \multicolumn{2}{c}{\textbf{No fusion}} & \multicolumn{2}{c}{\textbf{Low-level fusion}} & \multicolumn{2}{c}{\textbf{High-level fusion}} \\
    \cmidrule(lr){2-3} \cmidrule(lr){4-5} \cmidrule(lr){6-7}
    & L & R & Sum & Diff & Intersect & Union \\
    \hhline{-------}
    \textit{Moving average smoothing}\\
     \hspace{0.5cm}Single sensor window (sec) & 0.03 & 0.03 & 0.18 & 0.02 & 0.02 & 0.40 \\
     \hspace{0.5cm}Fused signal window (sec) & - & - & 0.08 & 0.02 & - & - \\
         \textit{Peak detection}\\
     \hspace{0.5cm}Min peak amplitude (range 0-1) & 0.27 & 0.36 & 0.15 & 0.08 & 0.24 & 0.06 \\
          \hspace{0.5cm}Min window between peaks (sec) & 0.36 & 0.34 & 0.23 & 0.40 & 0.34 & 0.32 \\
     \textit{Event fusion}\\
     \hspace{0.5cm}Max window for peak fusion (sec) & - & - & - & - & 0.32 & - \\
     \hspace{0.5cm}Min window between peaks (sec) & - & - & - & - & - & 0.29 \\
    \bottomrule
  \end{tabular}
\end{table}

The six step detection approaches are listed below. We denote $N_L$ and $N_R$ the signal magnitude (norm) of left and right sensors after smoothing, respectively. We denote $T_L$ and $T_R$ the set of times that match the peaks of $N_L$ and $N_R$, respectively, as identified by peak detection.

\textbf{No fusion} (two approaches, one per side): Peaks from only one sensor were used, either $T_R$ or $T_L$. 

\textbf{Low-level fusion} (two approaches): Fusion was performed at the raw data level after signal magnitude smoothing. Sum low-level fusion, $N_{R+L} = MovingAvg(N_R+N_L)$, or difference low-level fusion, $N_{|R-L|} = MovingAvg(|N_R-N_L|)$, were followed by peak detection to detect steps. $T_{R+L}$ and $T_{|R-L|}$  are the sets of times that match the peaks of $N_{R+L}$ and $N_{|R-L|}$.
 
\textbf{High-level fusion} (two approaches, see algorithms \ref{alg:intersection}, \ref{alg:union}): High-level fusion consisted of signal smoothing, peak detection, and fusion (intersection or union) of detected peaks. Fusion parameters included the maximum allowable distance $max\_dist$ to be considered an intersection (upper bounded by the peak detection minimum window parameter), or the minimum allowable distance $min\_dist$ between adjacent peaks (union). In both cases, the peak with the highest amplitude was selected.
\\
\begin{minipage}[t]{0.54\linewidth}
\vspace{0pt} 
\begin{algorithm}[H] 
\label{alg:Intersection}
\caption{High-Level Intersection}
\begin{flushleft}
\textbf{Input: $N_R, \: N_L, \: T_R, \: T_L$ and $max\_dist$} \\
\textbf{Output: $T_{R \cap L}$} (set of times)
\end{flushleft}
\begin{algorithmic}[1]
\label{alg:intersection}
    \STATE{$T_{R \cap L} = \emptyset$}
    \FOR {$t_l \in T_L$}
      \IF{$\exists t \in T_R: |t_l-t| \leq max\_dist$}
        \STATE{$t_r = argmin_{t \in T_R}(|t_l - t|)$}
        \IF{$\forall t \in T_L \quad |t_l-t_r|<|t-t_r|$}
         \STATE{$t_{r \cap l} = argmax_{t_r, \: t_l}(\{N_R(t_r), \: N_L(t_l)\})$}
         \STATE{$T_{R \cap L} = T_{R \cap L} \cup t_{r\cap l}$}
        \ENDIF
      \ENDIF
    \ENDFOR
\STATE{return $T_{R \cap L}$}
\end{algorithmic}
\end{algorithm}
\end{minipage}
\hfill\begin{tabular}[t]{p{.43\linewidth}}
\begin{minipage}[t]{\linewidth}
\vspace{0pt} 
\begin{algorithm}[H] 
\label{alg:Union}
\caption{High-Level Union}
\begin{flushleft}
\textbf{Input: $N_R, \: N_L, \: T_R, \: T_L$ and $min\_dist$} \\
\textbf{Output: $T_{R \cup L}$} (set of times)
\end{flushleft}
\begin{algorithmic}[1]
\label{alg:union}
    \STATE{ $S = T_R \cup T_L$}
	\STATE{ $T_{R \cup L} = \emptyset$}
    \WHILE  {$S \neq \emptyset$}
    \STATE{$P_{R \cup L} = N_R(T_R \cap S ) \cup N_L(T_L \cap S)$}
    \STATE{$p_i = max(P_{R \cup L})$}
    \STATE{$T_{R \cup L} = T_{R \cup L} \cup t_i$}
    \STATE{$S = S/\{t: |t-t_i| \leq min\_dist\}$}
    \ENDWHILE
    \STATE{return $T_{R \cup L}$}
    \newline
\end{algorithmic}
\end{algorithm}
\end{minipage}	
\end{tabular}

\section{Results}
\subsection{Step count accuracy}
\begin{figure}[ht]
  \centering
  \includegraphics[height=3.55cm,keepaspectratio,valign=t]{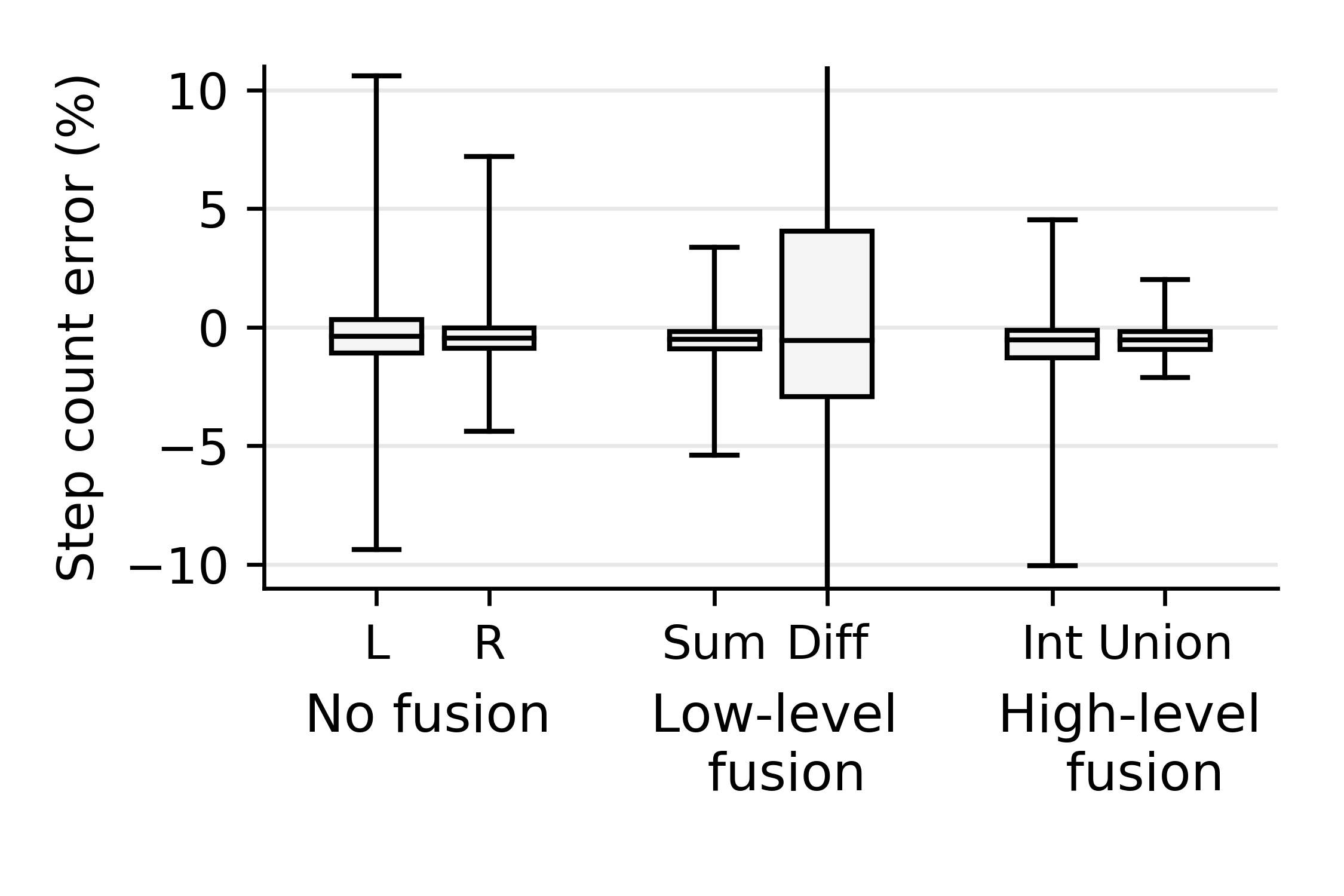}
  \hfill
  \includegraphics[height=3.25cm,keepaspectratio,valign=t]{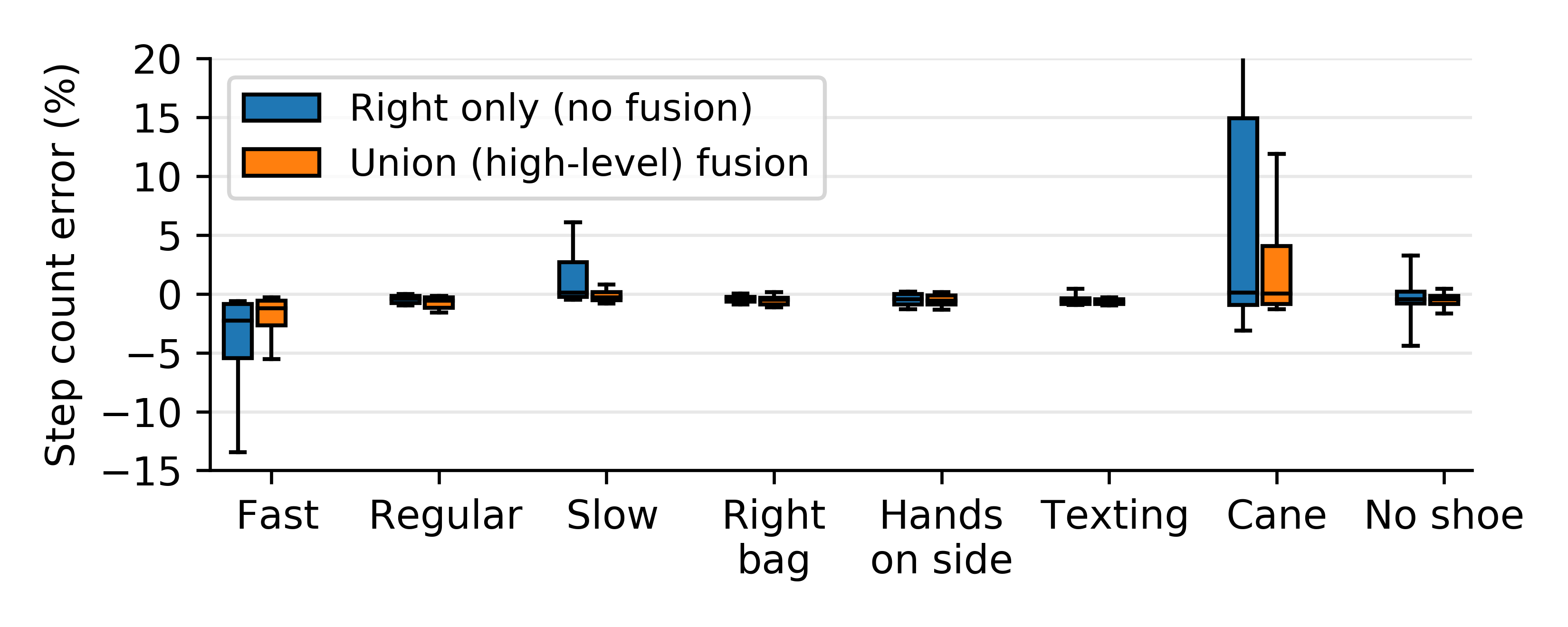}
  \caption{Union fusion has the lowest step counting error among sensor fusion approaches. Boxplot whiskers capture 90\% of data (5th and 95th percentiles). Negative error is under counting. Left: Overall step count error. Right: Comparison of union (high-level) fusion with the better no fusion algorithm (only right sensor) per task shows the union approach is robust to diverse types of walking.}
\label{fig:step_count_comp}
\end{figure}

Each one of the six algorithms were each trained using five-fold cross validation. Training was not per-task, but rather used all samples in order to evaluate algorithm robustness to different individuals and walking types. We observed that signal summation (low-level fusion) or union of peaks (high-level fusion) were better at quantifying step abundance than single side algorithms as measured by the percent error (Figure \ref{fig:step_count_comp}a). Union fusion also had substantially higher correlation to the label (ankle step count) than single sensor methods (union, r=0.98; left sensor, r=0.78; right sensor, r=0.88), with 90\% of the samples having a small step count error rate between -2.2\% and 2.5\% (see boxplot whiskers). Raw data fusion using difference of signal magnitudes performed poorly. 

Specific walking type comparisons of the better no fusion algorithm (right side) with the best fusion approach (high-level union) highlighted the potential for robust step counting using fusion (Figure \ref{fig:step_count_comp}b). While both methods performed well in multiple walking tasks, particularly where at least one arm was constrained, the no fusion algorithm was less accurate in fast or slow walking and in simulations of pathological gait (cane). In contrast, the robustness of union high-level fusion is strikingly evident.

\subsection{Gait phase detection}
\begin{figure}[ht]
  \centering
    \includegraphics[height=3.27cm,keepaspectratio,valign=t]{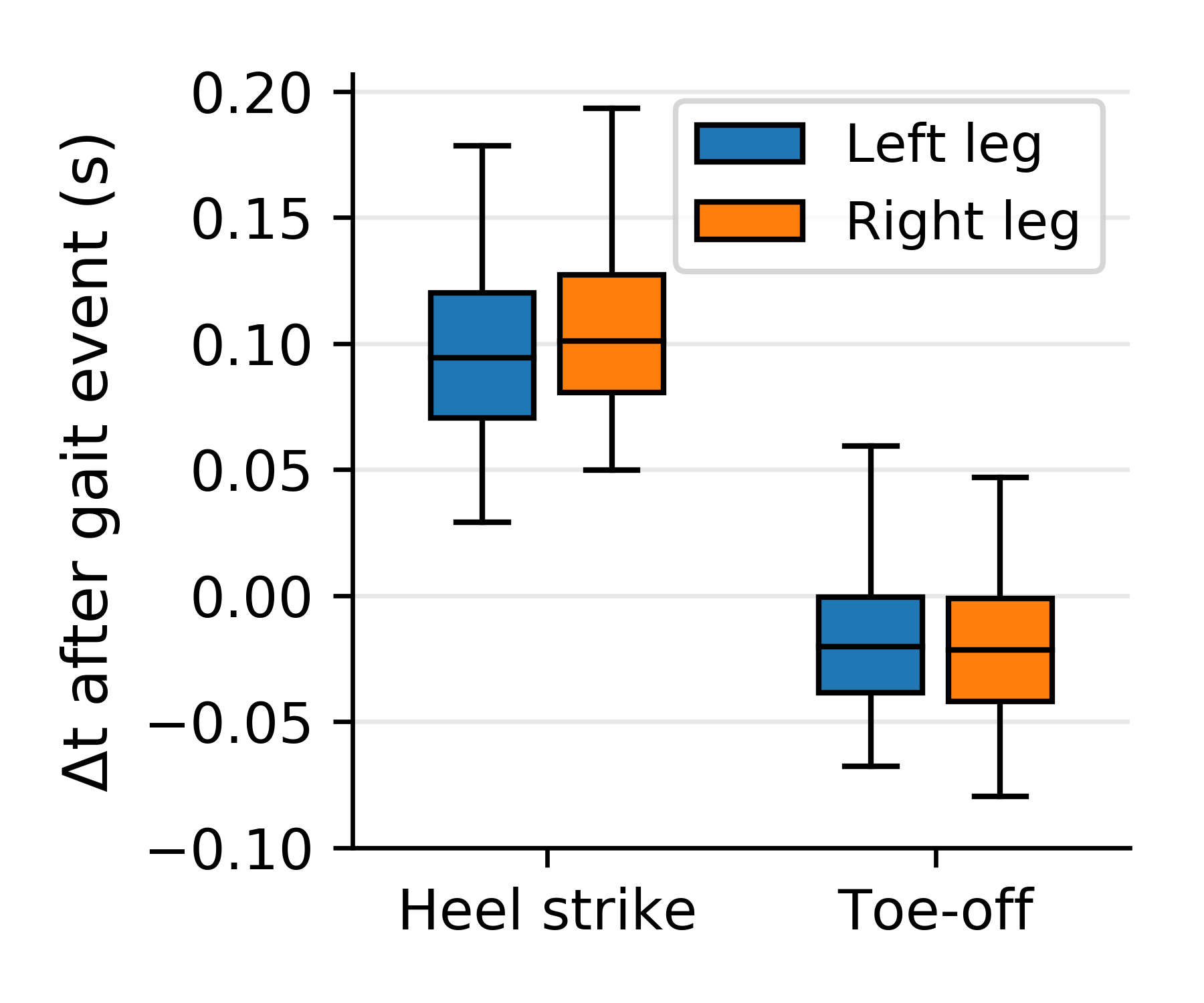}
  \qquad
  \includegraphics[height=3.5cm,keepaspectratio,valign=t]{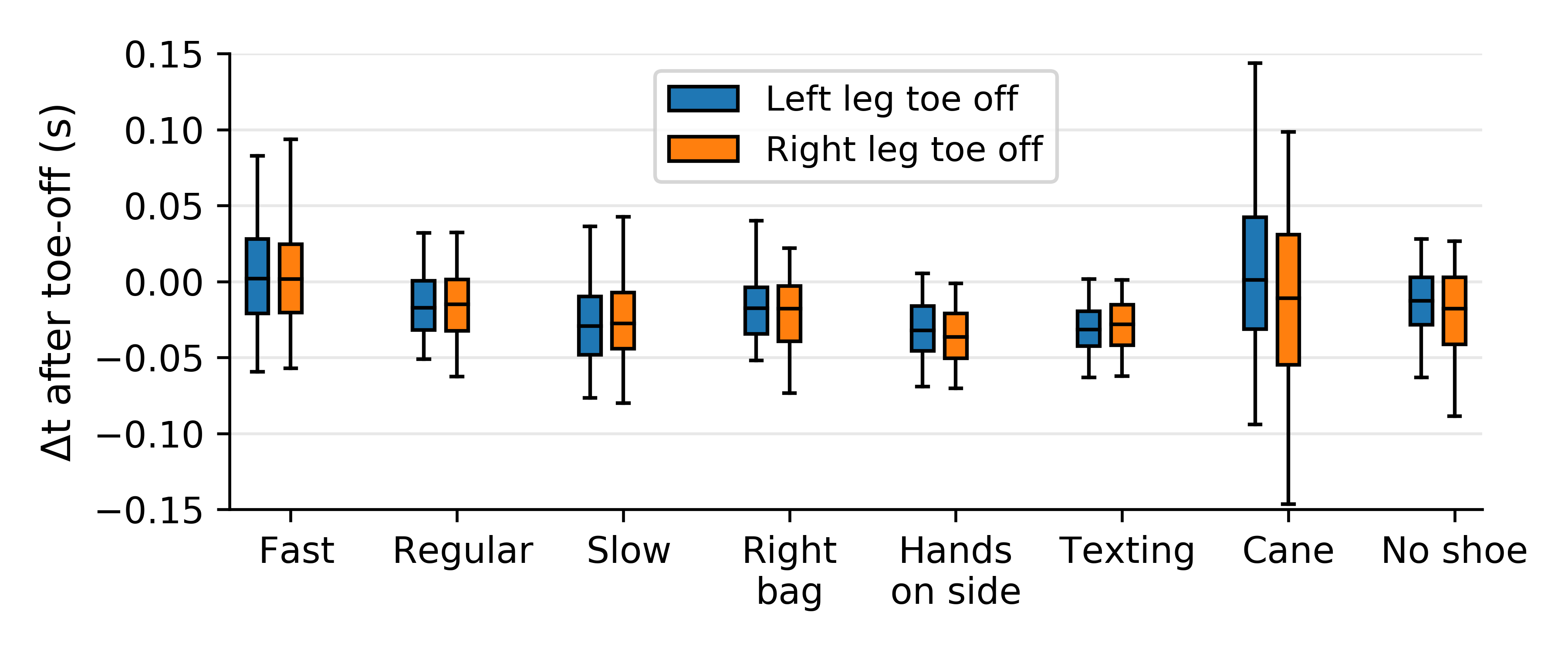}
  \caption{Gait phase detection by the peaks of the union (high-level) fusion approach. Left: Offset from heel strike and toe-off for all samples. Right: Offset from toe-off per walking task.
}
\label{fig:step_detection}

\end{figure}

The gait phase detected by all methods was expectedly similar as all methods share the same peak detection technique (data not shown due to space limitations). Focusing specifically on union (high-level) fusion, we observed that peaks were identified relatively consistently adjacent to the toe-off phase (Figure \ref{fig:step_detection}). The low variation in $\Delta$t offset relative to toe-off suggests that wrist-device fusion may enable quantification of gait asymmetries and cadence inconsistencies to a certain degree.

\section{Conclusions}
Our work has two primary contributions. First, we illustrate that mathematically simple fusion of wrist accelerometer output can overcome the shortcomings of single-wrist step counters. Further work is required to assess whether this holds in free-living environments. The second major implication of our work is the potential for dual wrist-devices to capture timing-based, gait abnormalities. As above, additional work can help evaluate this promise in patient populations.

\subsubsection*{Acknowledgments}
We thank all volunteers that participated in the study. We also thank Amitai Armon, Or Shimshi, Nir Darshan, Jeremie Dreyfuss, Gilad Wallach, and Abraham Israeli for insightful discussions.

\bibliographystyle{unsrtnat}
\bibliography{lib.bib}

\begin{thebibliography}{17}
\providecommand{\natexlab}[1]{#1}
\providecommand{\url}[1]{\texttt{#1}}
\expandafter\ifx\csname urlstyle\endcsname\relax
  \providecommand{\doi}[1]{doi: #1}\else
  \providecommand{\doi}{doi: \begingroup \urlstyle{rm}\Url}\fi

\bibitem[Richardson et~al.(2008)Richardson, Newton, Abraham, Sen, Jimbo, and
  Swartz]{richardson2008meta}
Caroline~R Richardson, Tiffany~L Newton, Jobby~J Abraham, Ananda Sen, Masahito
  Jimbo, and Ann~M Swartz.
\newblock A meta-analysis of pedometer-based walking interventions and weight
  loss.
\newblock \emph{The Annals of Family Medicine}, 6\penalty0 (1):\penalty0
  69--77, 2008.

\bibitem[Bassett et~al.(2017)Bassett, Toth, LaMunion, and
  Crouter]{bassett2017step}
David~R Bassett, Lindsay~P Toth, Samuel~R LaMunion, and Scott~E Crouter.
\newblock Step counting: A review of measurement considerations and
  health-related applications.
\newblock \emph{Sports Medicine}, 47\penalty0 (7):\penalty0 1303--1315, 2017.

\bibitem[Storm et~al.(2015)Storm, Heller, and Mazz{\`a}]{storm2015step}
Fabio~A Storm, Ben~W Heller, and Claudia Mazz{\`a}.
\newblock Step detection and activity recognition accuracy of seven physical
  activity monitors.
\newblock \emph{PloS one}, 10\penalty0 (3):\penalty0 e0118723, 2015.

\bibitem[Husted and Llewellyn(2017)]{husted2017accuracy}
Hannah~M Husted and Tamra~L Llewellyn.
\newblock The accuracy of pedometers in measuring walking steps on a treadmill
  in college students.
\newblock \emph{International Journal of Exercise Science}, 10\penalty0
  (1):\penalty0 146, 2017.

\bibitem[Chow et~al.(2017)Chow, Thom, Wewege, Ward, and
  Parmenter]{chow2017accuracy}
Jessica~J Chow, Jeanette~M Thom, Michael~A Wewege, Rachel~E Ward, and Belinda~J
  Parmenter.
\newblock Accuracy of step count measured by physical activity monitors: The
  effect of gait speed and anatomical placement site.
\newblock \emph{Gait \& Posture}, 57:\penalty0 199--203, 2017.

\bibitem[Chen et~al.(2016)Chen, Kuo, Pellegrini, and Hsu]{chen2016accuracy}
Ming-De Chen, Chang-Chih Kuo, Christine~A Pellegrini, and Miao-Ju Hsu.
\newblock Accuracy of wristband activity monitors during ambulation and
  activities.
\newblock \emph{Medicine and science in sports and exercise}, 48\penalty0
  (10):\penalty0 1942--1949, 2016.

\bibitem[Tudor-Locke et~al.(2015)Tudor-Locke, Barreira, and
  Schuna~Jr]{tudor2015comparison}
Catrine Tudor-Locke, Tiago~V Barreira, and John~M Schuna~Jr.
\newblock Comparison of step outputs for waist and wrist accelerometer
  attachment sites.
\newblock \emph{Medicine and science in sports and exercise}, 47\penalty0
  (4):\penalty0 839--842, 2015.

\bibitem[Del~Din et~al.(2016)Del~Din, Godfrey, Galna, Lord, and
  Rochester]{del2016free}
Silvia Del~Din, Alan Godfrey, Brook Galna, Sue Lord, and Lynn Rochester.
\newblock Free-living gait characteristics in ageing and parkinson’s disease:
  impact of environment and ambulatory bout length.
\newblock \emph{Journal of neuroengineering and rehabilitation}, 13\penalty0
  (1):\penalty0 46, 2016.

\bibitem[Mirelman et~al.(2016)Mirelman, Bernad-Elazari, Thaler, Giladi-Yacobi,
  Gurevich, Gana-Weisz, Saunders-Pullman, Raymond, Doan, Bressman,
  et~al.]{mirelman2016arm}
Anat Mirelman, Hagar Bernad-Elazari, Avner Thaler, Eytan Giladi-Yacobi, Tanya
  Gurevich, Mali Gana-Weisz, Rachel Saunders-Pullman, Deborah Raymond, Nancy
  Doan, Susan~B Bressman, et~al.
\newblock Arm swing as a potential new prodromal marker of parkinson's disease.
\newblock \emph{Movement Disorders}, 31\penalty0 (10):\penalty0 1527--1534,
  2016.

\bibitem[Elmenreich(2002)]{elmenreich2002sensor}
Wilfried Elmenreich.
\newblock Sensor fusion in time-triggered systems.
\newblock 2002.

\bibitem[Gao et~al.(2014)Gao, Bourke, and Nelson]{gao2014evaluation}
Lei Gao, AK~Bourke, and John Nelson.
\newblock Evaluation of accelerometer based multi-sensor versus single-sensor
  activity recognition systems.
\newblock \emph{Medical engineering \& physics}, 36\penalty0 (6):\penalty0
  779--785, 2014.

\bibitem[L{\'o}pez et~al.(2017)L{\'o}pez, Ocampo, Sucerquia, and
  Vargas-Bonilla]{lopez2017analyzing}
JD~L{\'o}pez, C~Ocampo, A~Sucerquia, and JF~Vargas-Bonilla.
\newblock Analyzing multiple accelerometer configurations to detect falls and
  motion.
\newblock In \emph{VII Latin American Congress on Biomedical Engineering CLAIB
  2016, Bucaramanga, Santander, Colombia, October 26th-28th, 2016}, pages
  169--172. Springer, 2017.

\bibitem[Fortune et~al.(2015)Fortune, Lugade, Amin, and
  Kaufman]{fortune2015step}
E~Fortune, VA~Lugade, Shreyasee Amin, and Kenton~R Kaufman.
\newblock Step detection using multi-versus single tri-axial
  accelerometer-based systems.
\newblock \emph{Physiological measurement}, 36\penalty0 (12):\penalty0 2519,
  2015.

\bibitem[Salarian et~al.(2004)Salarian, Russmann, Vingerhoets, Dehollain,
  Blanc, Burkhard, and Aminian]{salarian2004gait}
Arash Salarian, Heike Russmann, Fran{\c{c}}ois~JG Vingerhoets, Catherine
  Dehollain, Yves Blanc, Pierre~R Burkhard, and Kamiar Aminian.
\newblock Gait assessment in parkinson's disease: toward an ambulatory system
  for long-term monitoring.
\newblock \emph{IEEE transactions on biomedical engineering}, 51\penalty0
  (8):\penalty0 1434--1443, 2004.

\bibitem[Aminian et~al.(2002)Aminian, Najafi, B{\"u}la, Leyvraz, and
  Robert]{aminian2002spatio}
Kamiar Aminian, B~Najafi, C~B{\"u}la, P-F Leyvraz, and Ph~Robert.
\newblock Spatio-temporal parameters of gait measured by an ambulatory system
  using miniature gyroscopes.
\newblock \emph{Journal of biomechanics}, 35\penalty0 (5):\penalty0 689--699,
  2002.

\bibitem[Brajdic and Harle(2013)]{brajdic2013walk}
Agata Brajdic and Robert Harle.
\newblock Walk detection and step counting on unconstrained smartphones.
\newblock In \emph{Proceedings of the 2013 ACM international joint conference
  on Pervasive and ubiquitous computing}, pages 225--234. ACM, 2013.

\bibitem[Negri(2017)]{peakutils}
Lucas~H Negri.
\newblock \emph{PeakUtils 1.1.0}, 2017.
\newblock URL \url{https://pypi.python.org/pypi/PeakUtils}.

\end{thebibliography}
\end{document}